\documentclass[sigconf]{acmart}

\AtBeginDocument{%
  \providecommand\BibTeX{{%
    \normalfont B\kern-0.5em{\scshape i\kern-0.25em b}\kern-0.8em\TeX}}}

\copyrightyear{2024}
\acmYear{2024}
\setcopyright{rightsretained}
\acmConference[CHI EA '24]{Extended Abstracts of the CHI Conference on Human Factors in Computing Systems}{May 11--16, 2024}{Honolulu, HI, USA}
\acmBooktitle{Extended Abstracts of the CHI Conference on Human Factors in Computing Systems (CHI EA '24), May 11--16, 2024, Honolulu, HI, USA}
\acmDOI{10.1145/3613905.3650977}
\acmISBN{979-8-4007-0331-7/24/05}

\usepackage{wrapfig}

\begin{document}

\title[Digital Dotted Lines: Design and Evaluation of a Prototype for Digitally Signing Documents Using Identity Wallets]{Digital Dotted Lines: Design and Evaluation of a Prototype for Digitally Signing Documents Using Identity Wallets}

\author{Yorick Last}
\affiliation{%
   \institution{Paderborn University}
   \city{Paderborn}
   \country{Germany}}
\email{yorick.last@uni-paderborn.de}

\author{Jorrit Geels}
\affiliation{%
   \institution{Radboud University}
   \city{Nijmegen}
   \country{The Netherlands}}
\email{jorrit.geels@ru.nl}

\author{Hanna Schraffenberger}
\affiliation{%
   \institution{Radboud University}
   \city{Nijmegen}
   \country{The Netherlands}}
\email{hanna.schraffenberger@ru.nl}

\begin{abstract}
Documents are largely stored and shared digitally. Yet, digital documents are still commonly signed using (copies of) handwritten signatures, which are sensitive to fraud. Though secure, cryptography-based signature solutions exist, they are hardly used due to usability issues. This paper proposes to use digital identity wallets for securely and intuitively signing digital documents with verified personal data. Using expert feedback, we implemented this vision in an interactive prototype. The prototype was assessed in a moderated usability test ($N = 15$) and a subsequent unmoderated remote usability test ($N = 99$). While participants generally expressed satisfaction with the system, they also misunderstood how to interpret the signature information displayed by the prototype. Specifically, signed documents were also trusted when the document was signed with \textit{irrelevant} personal data of the signer. We conclude that such \textit{unwarranted trust} forms a threat to usable digital signatures and requires attention by the usable security community.
\end{abstract}

\ccsdesc[500]{Human-centered computing~Usability testing}
\ccsdesc[300]{Security and privacy~Key management}
\ccsdesc[500]{Security and privacy~Digital signatures}
\ccsdesc[100]{Human-centered computing~Heuristic evaluations}

\keywords{Digital Identity Wallets, Prototyping, Trust}

\maketitle

\section{Introduction}\label{sec:introduction}
It is increasingly common for documents to be signed on computers~\cite{Tzelios2020Psychological}.
Signing on a computer often involves scanning or drawing a digital version of a physical signature or even inserting a typed-out name (in a handwriting or script font), as offered in the popular signing tool Adobe Acrobat Reader~\cite{adobe_2023}. These kinds of signatures are referred to as ``electronic signatures''~\cite{eidas}. Signing documents this way is similar to physically signing documents and, therefore, intuitive for users~\cite{marcus1998metaphor,norman1986cognitive}. However, electronic signatures cannot reliably confirm the signer's identity or commitment~\cite{eu_esig}, regardless of their form. Namely, they can be easily copied, leading to potential misuse and fraud. Governments and businesses, therefore, regularly continue using physical signatures to be certain about the identity of the signatory~\cite{dijkhuis2018willeke} or employ proprietary solutions such as Adobe Acrobat Sign~\cite{adobe_sign_2022} or DocuSign~\cite{docusign_2023}. These digital signing solutions are generally closed-source, require data to be uploaded to external servers, and, above all, are costly. While their costs might be bearable for organizations, it is unreasonable to expect individuals to bear these costs. We are thus in need of secure, free, usable and trustworthy tools for digitally signing documents that can be used by organizations and \textit{individuals} alike. This paper sets out to develop such a signature tool and presents a first interactive prototype and its evaluation.

Cryptography offers secure and reliable digital signatures~\cite{merkle1987digital,merkle1989certified,bellare1999forward}. These are rooted in public-key cryptography~\cite{diffie1976new}, and provide guarantees over the signer's identity (``\emph{source authenticity}’’) and warrant that a message has not been changed (``\emph{message integrity}’’).

This cryptographic approach is already implemented in many tools for digitally signing email, and part of, e.g., PGP \cite{rfc4880-pgp}. However, in the context of signing (and encrypting) emails, the HCI community has identified many usability and adoption problems with such solutions ~\cite{Whitten1999johnny, garfinkel2005johnny, ruoti2019johnny, ruoti2015johnny, lerner2017confidante, garfinkel2005johnny}, in particular issues related to \emph{usable key management}~\cite{ruoti2019johnny}. It is extremely cumbersome to obtain and store cryptographic keys ~\cite{ruoti2015johnny, lerner2017confidante}, to verify the identity of key owners ~\cite{ulrich2011investigating, garfinkel2005johnny}, and to have your own identity verified~\cite{garfinkel2005johnny}. Furthermore, users seem to have a poor understanding of general cryptographic concepts (see, e.g.,~\cite{zissis2012cryptographic, abu-salma2017obstacles}), and many are unable to explain what digital signatures are~\cite{abu-salma2017obstacles,zaaba2015examination}. The lack of understanding of digital signatures may hinder the adoption of digital signature tools~\cite{abu-salma2017obstacles}, and even lead to security issues~\cite{lax2015digitaldocsigning}. 

Still, if the usability issues around key management are solved, cryptography-based signature applications could potentially provide secure, trustworthy, and usable signatures. A promising emerging solution to key management is \emph{digital identity wallets}. Identity wallets are applications that contain personal data of a user --- e.g., name, date of birth, or diplomas, verified by a \textit{trusted} issuing party, such as a governmental organization~\cite{whatidentwallet}. Users can then disclose these credentials to requesting parties to prove properties about themselves, i.e., who (e.g., a verified name) or what (e.g., a verified diploma) they are~\cite{hampiholi2015towards}. Digital identity wallets will become widely available in Europe soon~\cite{eu-wallet}, with current pilots investigating their use as a mobile driver's license, to open a bank account, or to collect prescriptions at pharmacies~\cite{eu_pilot}. Identity wallets can be used to \emph{replace complex cryptographic key management with the more familiar process of authenticating oneself}~\cite{DBLP:conf/chi/BotrosB0OSV23}: someone could sign a document with their name by proving who they are with the identity wallet. This process is likely more intuitive for both the signatory and the recipient. Namely, signatories interact with a QR code, a well-known interaction~\cite[see, e.g.,][]{ozkaya2015factors} whereas recipients can recognize signatories by their credentials rather than cryptographic keys~\cite[see, e.g.,][]{tosi2015digital}. 

Given the potential synergy between identity wallets and digital signatures, this paper explores how identity wallets can be used to provide a secure, usable, and trustworthy solution for digital signatures on documents and to what extent users understand the solution and the signatures they create. While digital identity wallets are becoming a popular topic among usability researchers~\cite{korir2022, DBLP:books/sp/22/KhayretdinovaKSR22, DBLP:conf/chi/BotrosB0OSV23}, wallet-based \textit{signing} has been explored only with a focus on cryptographic architecture~\cite{hampiholi2015towards} and as a potential means for battling fake news~\cite{bart}, leaving out usability aspects. In fact, digital \textit{document} signatures, in general, have received very little attention from the HCI community so far. Exceptions are the work by~\citet{math11020430}, who investigated various digital signature tools and their usability, and research by~\citet{fitriana2022design}, who have prototyped and tested a digital signature application for document legalization. However, as these studies focus on different signing mechanisms (e.g., logging into a platform with a password to sign), the potential of digital identity wallets for signing still remains unexplored.

To investigate this potential, we created \emph{IdentitySign} --- an interactive prototype for digitally signing PDFs with the identity wallet \textit{Yivi}. This wallet is a front-runner of the European Digital Identity Wallet and is already actively used in the Netherlands (with $>$150k registrations). The prototype was developed using expert feedback, subsequently evaluated and improved using a moderated usability test, and finally assessed with a large-scale unmoderated remote usability test.

Our results show that although participants were satisfied with wallet-based signing, they generally misunderstood the role and meaning of digital signatures. Specifically, participants trusted signatures even when this was not warranted (e.g., because the signature was placed by the wrong entity). Another finding is that the role of the digital identity wallet in the signing process was also often unclear to participants. These findings contribute to the highly relevant domain of both digital signatures and digital identity wallet apps. 

\section{The IdentitySign prototype}\label{sec:prototype}
Our prototype for digitally signing PDF documents is called \emph{IdentitySign} and takes the form of a JavaScript/TypeScript web application that runs client-side within the user’s browser. The source code of \emph{IdentitySign} is available in the supplemental materials, and screenshots of all features are available in Appendix \ref{app:a}. The prototype works with the identity wallet \textit{Yivi}. In the Netherlands, \textit{Yivi} can be filled with verified data from official issuers (e.g., a municipality). These data can be used to sign documents. \emph{IdentitySign} also supports signing with demo credentials, to allow testing from outside the Netherlands and to facilitate anonymous participation in our usability studies.

We developed \emph{IdentitySign} iteratively, moving from low-fidelity mockups to an interactive website with a fully functional front-end. An early version of \emph{IdentitySign} was improved with the help of feedback from four UX experts, who assessed the usability using the 10 usability heuristics by Nielsen~\cite{nielsen-heuristics}, identified UI legibility and consistency issues and suggested improvements, such as providing more explanation to users. The final prototype features three core functionalities, namely, (1) signing documents, (2) verifying signatures, and (3) creating signature requests.  Because we were interested in the prototype's usability and users' understanding of the signatures, we implemented only the front-end of the application. \emph{IdentitySign} does not (yet) make use of cryptography to create actual digital signatures. (Actual cryptographic operations are not relevant to our specific research as they happen `under the hood' rather than affect the user flow/interface.) From a user's perspective, \emph{IdentitySign} appears to be fully functional. While all functionalities are provided on our \emph{IdentitySign} website, the entire process runs client-side, ensuring documents remain confidential and are not uploaded to some cloud service as with, e.g., DocuSign~\cite{docusign_2023}. 
\subsection{Signing documents}

\begin{figure*}[ht]
		\centering
		\includegraphics[width=\textwidth]{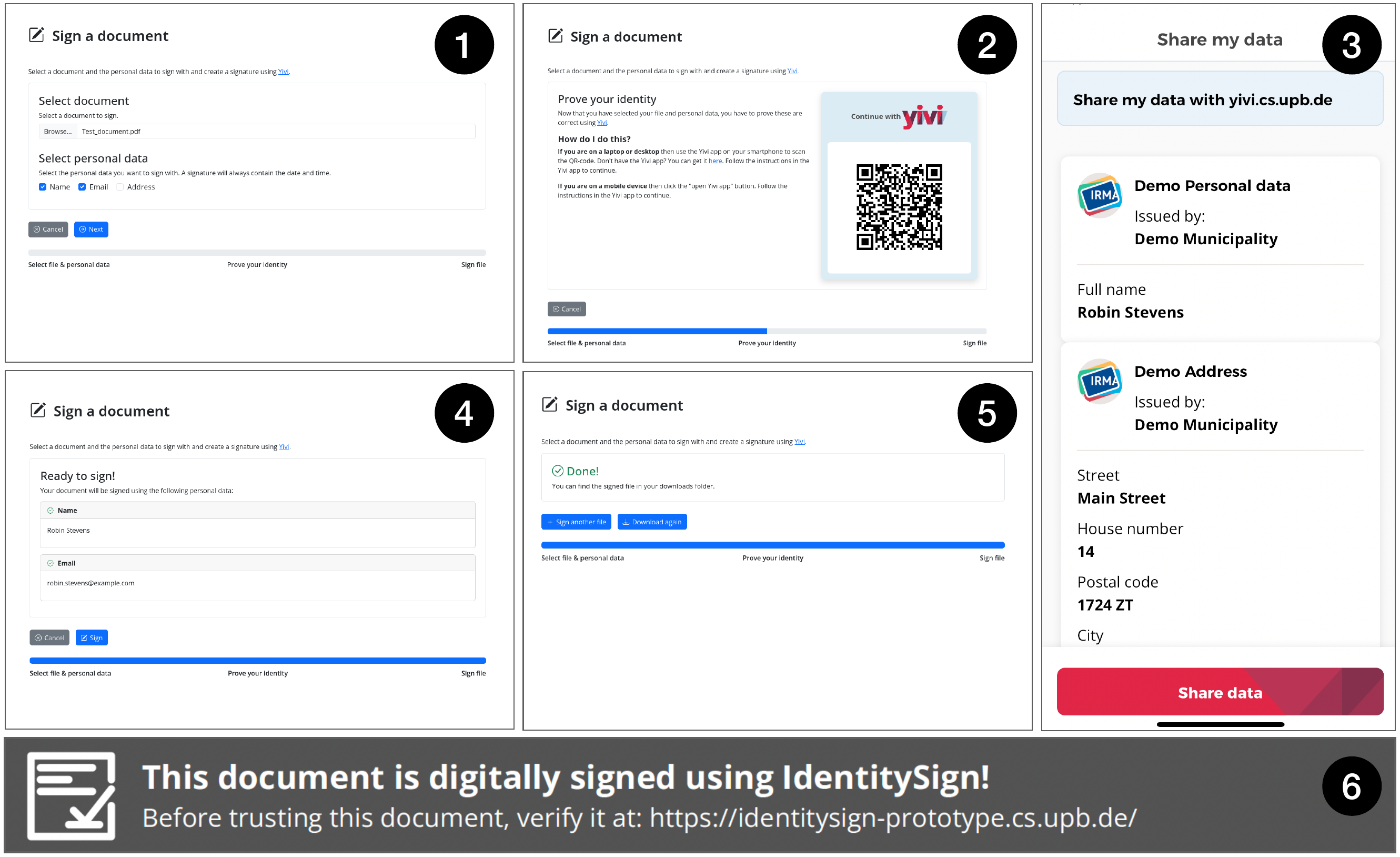}
		\caption{User flow of signing a PDF with \emph{IdentitySign} and \textit{Yivi}. Screens have been cropped to display the main elements.}
        \Description{Screenshots of each step in the process of signing a document (as described in the main text).}
        \label{fig:flow}
\end{figure*}

To create a signature, users have to select the document they wish to sign and the personal data they wish to sign with on the \emph{IdentitySign} website (see Fig. ~\ref{fig:flow}, step 1). For instance, a user might select a file ``\textit{document.pdf}'' and decide to sign with their name and email address. Subsequently, users need to scan a QR code with \textit{Yivi} (2), which prompts them to share the selected \textit{Yivi} credentials (containing name and email address) with the \textit{IdentitySign} application, thus proving who they are (3). Subsequently, the disclosed credentials show up in \textit{IdentitySign}, and the user can click ``Sign'' to sign the PDF (4). Upon doing so, the PDF is signed by the client-side software and can be found in the download folder (5). The signed PDF contains a footer at the bottom of the PDF that informs the recipient about the signature and about where to verify the signature (6).

\subsection{Verifying signatures}
\begin{figure*}[ht]
		\centering
	    \includegraphics[width=0.68\textwidth]{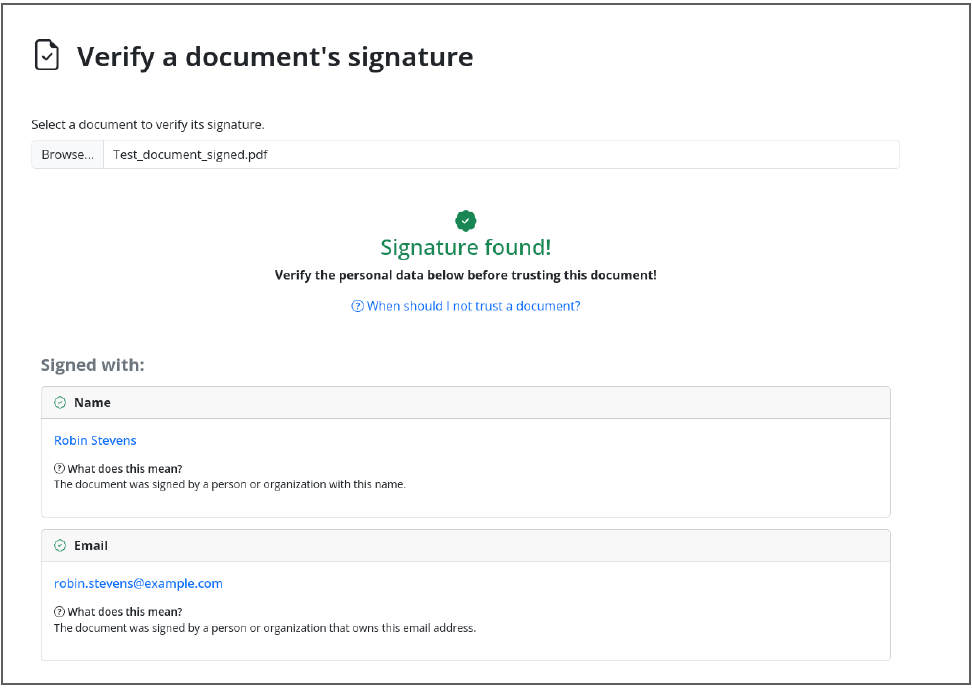}
		\caption{Verification screen with a valid signature (final implementation, cropped to show the main part of the screen).}
        \Description{Screenshot of the prototype's verification screen, displaying the filename of the selected file, a centered green checkmark with the text "signature found", and the name and email used to create the signature.}
        \label{fig:verify}
\end{figure*}

A signature can be verified by selecting a signed digital document within \emph{IdentitySign}.\footnote{Signing happens digitally, and \textit{IdentitySign}'s signatures on \textit{printed} documents cannot be verified.} If the document contains a signature, \emph{IdentitySign} will then display a ‘valid signature’ screen, which contains the personal data that was used to sign the document (see Fig.~\ref{fig:verify} for the final implementation). If the document does not contain a signature, a message informs the user of this.

\subsection{Requesting signatures}
Signatures can be requested by determining \textit{who} should sign which \textit{document} with which \textit{credentials}. This involves selecting the file that should be signed and the attributes (e.g., name, address) that the signer should sign with. \emph{IdentitySign} then generates a signature request link and an accompanying message, which can be copied with one click and subsequently be shared together with the file that needs to be signed, e.g., via email or text message. Opening such a link will restrict file selection to a file of the same name and will pre-select the personal data used for signing.

\subsection{Limitations and envisioned implementation}\label{sec:protolimit}
As mentioned, we have implemented only the front-end and user flow of the application, i.e., no actual (cryptographic) signatures are created yet. However, the \textit{technical} feasibility of using \textit{Yivi} to sign files digitally is demonstrated by the \textit{PostGuard} project~\cite{DBLP:conf/chi/BotrosB0OSV23}, where users prove who they are with \textit{Yivi} to encrypt/decrypt emails. Like \textit{PostGuard}, \textit{IdentitySign} can use identity-based encryption~\cite{shamir1984identity, bonehIdentityBasedEncryptionWeil2001, boneh2004efficient}, and ask users to authenticate themselves to a trusted third party, i.e., an \emph{IdentitySign} server, with the credentials they want to include in their signature to sign documents.
 
While we have implemented the flow with \textit{Yivi}, \textit{IdentitySign} could equally work with different (future) identity wallets as long as they allow users to authenticate themselves via selective disclosure of attributes (e.g., name, address).

\section{Evaluation of IdentitySign}
We evaluated \emph{IdentitySign}'s usability and users' trust in the provided digital signatures with two (IRB-approved) complementary, subsequent studies. Study 1 ($N = 15$) was a moderated usability test and took place in person, whereas study 2 ($N = 99$)  was unmoderated and carried out remotely about a month later. Both studies gathered qualitative and quantitative data. However, study 1 was more exploratory, focused more on qualitative aspects, used the ‘think-aloud’ method~\cite{lewis1993task, nielsen1994usability}, and allowed us to observe participants and ask clarifying questions. In contrast, study 2 focused more on obtaining quantitative data and potentially generalizable results. We kept methods and materials constant across studies as much as possible. Any differences are listed below.

\subsection{Participants}
To be eligible for participation in our studies, participants had to be at least 18 years of age, able to use a computer and smartphone, able to speak and read English, and they needed to be based in the European Union. Experience with digital identity wallets was not required.

For study 1, the moderated usability test, we recruited 15 participants from our own network. Of the 15 participants, 9 were between 18–35 years old (60\%), 3 between the age of 36 and 55 (20\%), and 3 over 55 years old (20\%). Most participants either completed or were currently attending higher education (\textit{66.7\%, N =10}), followed by vocational education (\textit{20\%, N =3}) and primary or secondary education (\textit{13.3\%, N =2}). Two of the participants (\textit{13.3\%}) either studied or were employed in a field related to IT.

We recruited 100 participants via Prolific for study 2, the large-scale unmoderated usability test. For the analysis, we excluded one participant who was unable to finish due to technical difficulties. Participants of the second study were slightly younger overall, with most being in the 18-35 age group (\textit{77.8\%, N =77}), followed by the 36-55 (\textit{17.2\%, N =17}) and the >55 (\textit{5.1\%, N =5}) age group. Most participants either completed or were currently attending higher education (\textit{76.8\%, N =76}), followed by primary or secondary education (\textit{17.2\%, N =17}) and vocational education (\textit{5.1\%, N =5}). Field of study/employment was not recorded for this study.

\subsection{Procedure}
After an informed consent procedure, participants completed a pre-study questionnaire
and received pre-task information about digital signatures, our prototype, and \textit{Yivi}. Study 1 continued with a short demo of the identity wallet \textit{Yivi}, upon which participants were handed a laptop running Firefox (for \textit{IdentitySign}) and Thunderbird (for receiving/sending documents) and a mobile phone running \textit{Yivi}~\cite{irma} (with a demo credential with the name Robin Stevens inside the wallet). 
In study 2, participants received instructions about installing \textit{Yivi} on their phones and loading the demo credential into \textit{Yivi}. 
Participants were then asked to complete a number of tasks (see below), each followed by a post-task questionnaire.
In the end, participants completed a post-study questionnaire.
Finally, participants were thanked and then either presented with a gift voucher (study 1) or returned to Prolific for payment (study 2).

\subsection{Tasks and materials}
In both studies, we asked participants to complete signature-related tasks. These tasks were presented in the form of short motivating scenarios. The exact materials used for both studies can be found in the supplemental materials.

\subsubsection{Study 1}
The moderated usability test included four tasks ($A_1$, $B_1$, $C_1$, $D_1$). \textit{Tasks \textbf{$A_1$}} was designed to evaluate the \textit{signing process}. To complete this task successfully, participants had to download a PDF and sign the document with specific personal data. \textit{Task \textbf{$B_1$}} meant to evaluate \textit{signature requests} from a recipients/signers perspective. To complete it, participants had to open a signature request received via mail, sign the attached document, and return it to the requestor. \textit{Task \textbf{$C_1$}} was created to evaluate \textit{signature requests} from a requestors' perspective. Participants needed to obtain a signature on a sales agreement by using the signature request feature. This entailed generating the request and sharing it by email, together with the document that needed to be signed. They then received the requested signed document, enabling us to see whether participants would verify the signature on the received document with \textit{IdentitySign}.  \textit{Task $D_1$} was included to evaluate the \textit{signature verification} process. Participants were told to verify the certificate (diploma) of an electrician and decide whether to hire the electrician. Notably, the signature on the electrician's certificate was placed by the electrician themselves instead of the institution that awarded the certification (thereby not providing any guarantees of the legitimacy of the diploma). We thus wanted to find out whether participants would consider \emph{who} signed the document and identify a meaningless signature as such (instead of trusting the document simply because of the presence of a signature).

As this first study primarily served an exploratory purpose, we made two slight adjustments to the test materials in between test sessions. Because initially, most participants did not trust the electrician's certificate in task $D_1$ due to the certificate's simplistic and unrealistic design, we switched to a realistic-looking certificate. This change seems to have mitigated the issue. As users were not alarmed by the fact that the electrician's certificate was signed by the electrician themselves, we furthermore included \textit{warnings} to not blindly trust signatures in the prototype and added help information on \textit{when not to trust a signature}. However, we did not observe a difference in the subsequent task results. 

After the first study, we implemented the functionality needed for setting up the \textit{Yivi} app with demo credentials. This was necessary as participants in study 2 would not have access to a smartphone with the \textit{Yivi} application pre-configured.

\subsubsection{Study 2}
The unmoderated remote usability test included two of the tasks from study 1, namely adapted versions of task $A_1$ and $D_1$, which we refer to as task $A_2$ and $D_2$. Accordingly, \textit{task $A_2$} focused on \textit{signing} a document and \textit{task $D_2$} on the \textit{signature verification} process. The tasks were changed slightly to fit an unmoderated remote study. For instance, in task $A_2$, a success code was added to correctly signed PDFs. (This code needed to be reported back to us, so we could check if participants had completed the task successfully despite the remote and unmoderated setting.) The \emph{core task} of downloading and signing a document (task $A_1$/$A_2$) and verifying a signature (task $D_1$/$D_2$) remained the same in both studies. (The exact scenarios of $A_1$ and $A_2$ differed, with participants signing a building permit in study 1 and a lease agreement for a storage box in study 2.) 

\subsection{Measures and data}
Both studies collected quantitative and qualitative data on usability, trust, and general impressions.

We collected demographics, participants' experience with signing digital documents and existing digital signature tools, and their knowledge and understanding of digital signatures. Whereas we included  Affinity for Technology Interaction (ATI) scale~\cite{ati} in study 1, it was excluded in study 2 for survey length reasons.

For each task, we measured task success and factors expected to affect perceived usability with the After-Scenario-Questionnaire (ASQ)~\cite{DBLP:journals/ijhci/Lewis95}. Overall usability was measured with the System Usability Scale (SUS)~\cite{brooke}. For these quantitative measures, descriptive statistics were computed.

Our questionnaires also included open questions (e.g., to record whether participants liked/disliked anything in particular) and asked participants to rate several aspects, e.g., trust in \textit{IdentitySign}'s digital signatures, via self-designed scales.

In study 1, participants were observed, and audio was recorded and later transcribed. 

\section{Results}
This section presents quantitative results and qualitative insights.

\subsection{Quantitative results}\label{sec:quant}
Quantitative data was evaluated for each study separately. We first present success rates and usability perceptions (ASQ ratings) per task, followed by results concerning the overall usability of the prototype (SUS). 

\subsubsection{Task A}
Downloading and signing a document was completed with a success rate of 11/15 (study 1; $A_1$) and 73/99 (study 2; $A_2$). Task $A_1$ received an ASQ score of 4.9 ($SD = 1.30$), and task $A_2$ a score of 5.65 ($SD = 1.46$) on a scale of 1 to 7, where higher scores represent higher task usability. As reflected by these results, most participants found this task relatively easy and did not encounter any major issues. However, some participants had issues in recognizing whether the signing process was completed (most common in study 1, probably due to the absence of the success code introduced in study 2) or ran into issues with setup (most common in study 2).

\subsubsection{Task B}
Responding to a signature request was only featured in study 1. The task ($B_1$) was completed successfully by 14/15 participants. It received an ASQ score of 5.3 ($SD = 0.93$). Some participants noted that they perceived this second task as easier because they were more familiar with how \emph{IdentitySign} worked after the first task.

\subsubsection{Task C}
Requesting, obtaining, and verifying a signature was only featured in study 1. This task ($C_1$) had a success rate of 9/15, and an ASQ score of 5.9 ($SD = 1.09$). Problems that caused task failure included not sending a document to sign, not utilizing the requested functionality, or not verifying the received document.

\subsubsection{Task D}
Verifying the signature of a diploma was completed successfully by 3/15 (in study 1; $D_1$) and 19/99 (in study 2; $D_2$). The ASQ scores were 5.6 ($SD = 1.27$) for $D_1$ and 6.08 ($SD = 1.22$) for $D_2$. In other words, most participants were not able to successfully complete this task. We found two common reasons of task failure. First, some participants did not check the signatory. Second, some participants checked the signatory, but failed to question whether the assurances of that signatory were sufficient for the document in question. In neither study, participants explicitly expressed confusion about the task or scenario itself. However, observing confusion for participants was difficult in study 2 due to the remote setup.

\subsubsection{Overall usability}
The overall prototype got a SUS score of 73.5 ($SD = 17.39$) in study 1 and 76.4 ($SD = 16.3$) in study 2. According to \citet{bangor2008}, these SUS scores can both be interpreted as ``good''.

\subsection{Qualitative results}\label{sec:qual}
\begin{figure*}[ht]
        \centering
        \includegraphics[width=0.8\textwidth]{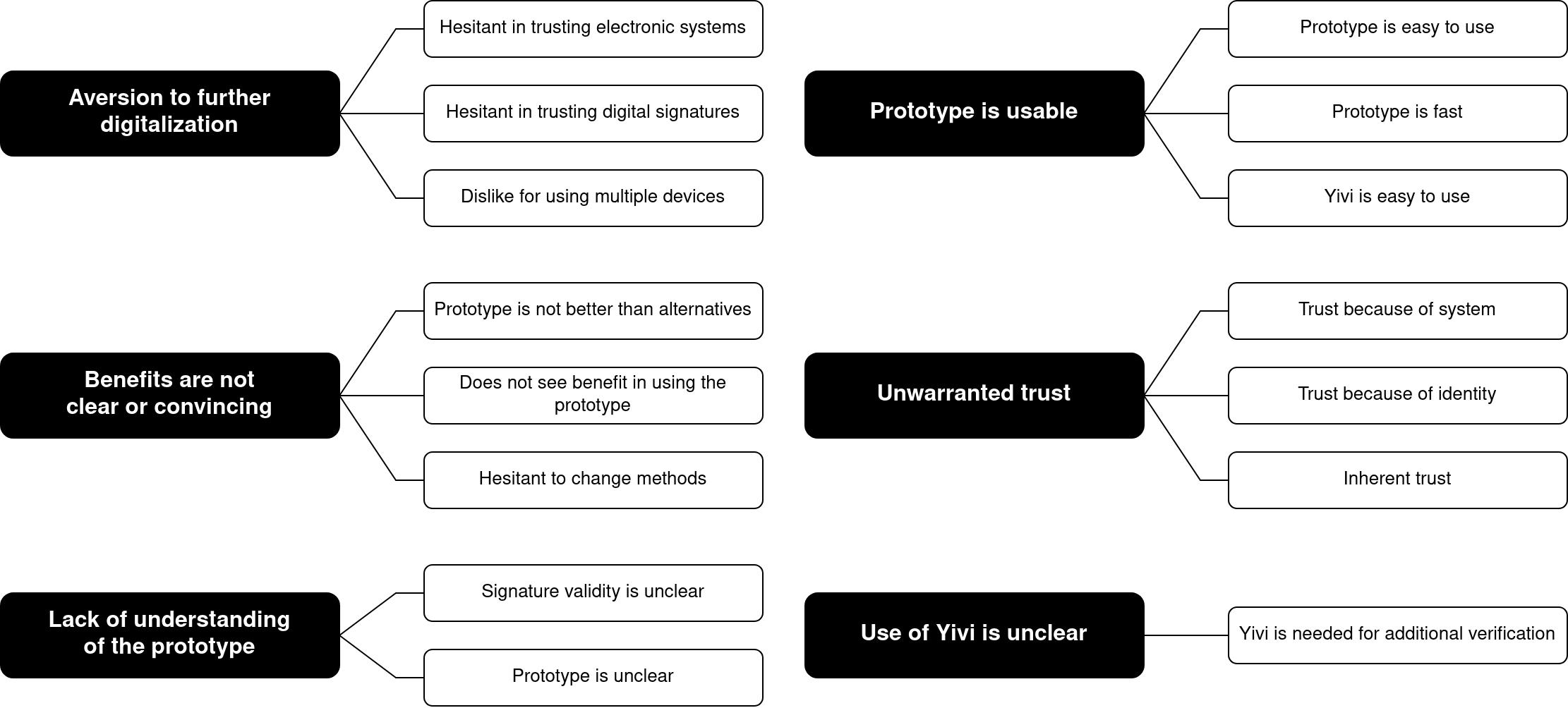}
        \caption{Visual representation of thematic analysis themes.}
        \Description{Diagram containing a visual breakdown of main themes, namely 1: "aversion to further digitalization" into "hesitant in trusting electronic systems", "hesitant in trusting digital signatures" and "dislike for using multiple devices", 2: "benefits are not clear or convincing" into "prototype is not better than alternatives", "does not see benefit in using the prototype", and "hesitant to change methods", 3: "lack of understanding of the prototype" into "signature validity is unclear" and "prototype is unclear", 4: "prototype is usable" into "prototype is easy to use", "prototype is fast", and "Yivi is easy to use", 5: "unwarranted trust" into "trust because of system", "prototype is fast", and "Yivi is easy to use", and 6: "use of Yivi is unclear" into "Yivi is needed for additional verification".}
        \label{app:themesdiagram}
    \end{figure*}

We conducted a thematic analysis \cite{braun_thematic_2012} of the combined qualitative data of both studies. In our coding, we marked participants' comments on our prototype, \textit{Yivi}, digital signatures in general, and their current method of signing documents. Positive and negative remarks on each of these topics were assigned separate codes. From these codes, themes were identified, which were subsequently reviewed and refined.
This section presents the resulting themes, followed by suggested improvements. Figure \ref{app:themesdiagram} provides a visual summary of the themes.

\subsubsection{Prototype is usable}

Participants were generally very positive about \emph{IdentitySign}'s overall usability. Positive reactions to the prototype were mostly related to participants describing the prototype as easy to use and fast. As one participant in study 2 stated: \textit{``All the information needed to use \textit{IdentitySign} was available on the front page and the process was very easy to follow.''} Similarly, positive reactions about \textit{Yivi} were mostly about its ease of use. As stated by another participant in study 2: \textit{``Yivi has a great UI and seems very straightforward.''}

\subsubsection{Unwarranted trust}
While participants were able to complete most tasks with \emph{IdentitySign} successfully, a big problem we observed was unwarranted trust. Participants commonly trusted the signatures made by the prototype, either because of trust in the prototype in general or trust in the information displayed with the signature. Participants' trust was, e.g., based on the visible banner on the PDF, motivated by the presence of a signature, or caused by the green checkmark and/or signer's name displayed in the verification process. This was most notable for tasks $D_1$ and $D_2$, where participants had to verify the certification of an electrician. Most participants incorrectly assumed the certificate's validity --- even though the certificate was signed by the electrician themselves rather than by the institution responsible for the certification. One participant in study 2 stated \textit{``I got a signature as a result, so the electrician was certified.''}

\subsubsection{Lack of understanding of the prototype}
Our qualitative data revealed that the \emph{IdentitySign} prototype was not entirely clear. Misunderstandings mostly arose from signature validity not being clear or the information contained within the prototype itself not being sufficiently clear. As stated by a participant in study 2: \textit{``It says the signature was found, but I am not sure what that means.''} Some participants also noted that the signature did not contain sufficient information for them to judge whether they could trust it.

\subsubsection{Use of Yivi is unclear}
The role of \textit{Yivi} in the signing process was also not always clear to participants, with some thinking that \textit{Yivi} served as an additional layer of verification/authentication. Two participants in study 1 described the role of \textit{Yivi} as \textit{``Yivi is the 2-step verification to confirm the identity of the person making the signature.''} and \textit{``Possibly for double identity check.''}. While\textit{Yivi} is indeed used to prove one's identity,  there is no \textit{second/double} check of the user's identity.

\subsubsection{Benefits are not clear or convincing}
We identified roadblocks in the way of adoption. Specifically, the benefits of using digital signatures, \emph{IdentitySign}, or both were either unclear or unconvincing to some participants. These participants were hesitant to change their methods and often preferred their current method of signing over the prototype. As one participant in study 1 states: \textit{``I’m a bit on the fence. The program is easy to use and, I assume, very reliable. But I don’t think I would choose this over Acrobat.''}

\subsubsection{Aversion to further digitalization}

On a more general level, we observed an aversion to further digitalization, which also could be an obstacle to adoption. Some participants were hesitant to trust either electronic systems in general or digital signatures specifically. One participant in study 2 notes: \textit{``Not sure I would want to use it in the real world though. Too much information is being stored digitally for my liking. Who would be able to see the information?''}, illustrating their reluctance to store personal information digitally. These participants were worried about the trustworthiness of electronic systems and/or preferred to stick to traditional, handwritten signatures. Some also mentioned a dislike for having to use another device, either because participants did not want to switch devices or carry a second device.

\subsubsection{Suggested improvements}
Multiple participants indicated their desire for a visual representation of a signature instead of the banner used by the current prototype. Two common reasonings behind this were 1) the banner not being recognizable as a \textit{signature}, and 2) the banner did not evoke the level of trust a signature would.

\section{Discussion}\label{sec:discuss}
We have presented a prototype for digitally signing documents with an identity wallet. Our main goal was to create an easy-to-use, efficient method to sign digital documents securely. Our prototype was largely well-received, and our results suggest that this is achievable with an identity wallet. In fact, for many users, the ease of use and speed with which they could sign and verify documents were reasons to be willing to use \emph{IdentitySign} in the future. Specifically, the signing-focused tasks  ($A_1$, $A_2$ and $B_1$) were successfully completed by most participants. However, our evaluation revealed serious issues, especially with signature verification ($D_1$ and $D_2$), which require further study and might also play a role in other signature apps.

The biggest issue we observed is that users trusted the signatures even in situations where this was not warranted, e.g., when a document was signed with irrelevant personal data. This unwarranted trust is likely rooted in users' habit of not scrutinizing the legitimacy of a signature until the assumption of honest communication is lifted~\cite{levine2014,levine2022}. If digital signatures can be used to mislead users, they are not secure in practice~\cite{cranor2005security}. 

Participants seemed to over-trust signatures because of checkmarks and green text outputted by the prototype, i.e., \textit{automation bias}~\cite{SKITKA1999991}, or because of the presence of a signature banner. Interestingly, a similar \textit{``inadequate reliance on visual cues as a proxy for proper digital verification''}~\cite{GerhardtPDK23} was also observed by~\citet{GerhardtPDK23} in the related context of visual digital certificates. Avoiding potentially misleading visual elements such as checkmarks and words such as ``valid'' could help, as users tended to interpret these as ``trustworthy''~\cite[see, e.g.,][]{hilligoss2008developing}. However, as~\citet{GerhardtPDK23} point out, using visual cues to determine authenticity is common with analog documents. Hence, adapting users' verification behavior might require additional effort~\cite{GerhardtPDK23}. As a next step, we plan a re-design that nudges users to be more conscientious when deciding whether to trust a signed document without harming overall usability by, for example, incorporating \emph{``security enhancing friction''}~\cite{distler2021}. 
		
Multiple participants indicated that they missed a visual representation of a handwritten signature. The absence thereof might negatively affect the perceived validity and trustworthiness of signatures made by \emph{IdentitySign}~\cite[see][]{chou2015Paperless}. Hence, future work could explore the effects of adding a visual representation of a signature, e.g., by displaying the attributes used for signing in a calligraphic manner. However, we worry that such visual additions might cause users to inadequately rely on these visual cues~\cite{GerhardtPDK23}. A key question for follow-up work is how to design the signature banner so it evokes associations with a signature but also \emph{prompts users to inspect and verify the actual (digital) signature}.

This research is one of the first to look into use cases of digital identity wallets. 
Overall, \textit{Yivi} and the interaction of \emph{IdentitySign} with Yivi were well-regarded by participants. Those who disliked the use of \textit{Yivi} mainly raised complaints about the need to use multiple devices and the reliance on QR codes. Still, this was also explicitly mentioned as a positive property by some. User experience spanning multiple devices comes with challenges that have been previously recognized and studied~\cite{brudy2019}. QR code alternatives, such as push notifications, could be explored in further studies.

Given the novelty of identity wallets, it is not surprising that the role of \textit{Yivi} in the signing process was not well understood, even though users generally used it correctly. The introduction of a European Digital Identity might help overcome such problems and lower the initial adoption effort. Namely, if users already have an identity wallet installed, it no longer becomes necessary to install an additional mobile application specifically for signing documents. 

Our concept assumes that the identity wallets themselves will provide users with a good user experience, something that has not been widely studied yet~\cite{korir2022, sellung2023}. More research into the usability of identity wallets is thus also desirable.

There are some limitations to our results. First, participants in study 1 might have reacted more positively as the moderator was also the developer of \emph{IdentitySign}, which is undesirable~\cite{rubin2008handbook, lazar2017research}. However, this effect seems to be marginal, given that study 2 produced similar results. 
Second, our design process has involved users rather late, possibly affecting our design solution space. In hindsight, users could have been involved earlier through, e.g., a participatory design~\cite{spinuzzi2005methodology} session. Third, as (some forms of) signatures hold legal value, it is desirable to include a legal perspective in our design process, too. Fourth, using \textit{Yivi} demo credentials might have influenced our results. Namely, filling the \textit{Yivi} app with one's own real credentials can take slightly more effort. Still, it is reasonable to assume that users will have a filled identity wallet readily at hand in the near future.
Fifth and last, it is unclear how our results generalize to future EU digital identity wallets, as their user experience might not necessarily be similar to \textit{Yivi}.

\section{Conclusion}
Our prototype introduces a novel use case of digital identity wallets to create digital signatures. Based on our evaluation, we conclude that the concept of digital signatures using identity wallets is promising but that the current prototype should not be published. Arguably the most important issue is the unwarranted trust that the mere \textit{presence} of a signature can induce. This poses the threat of digital signatures being used to legitimize fraudulent documents and has implications beyond merely the effectiveness and usability of this specific prototype.

We believe a solution to this problem exists, e.g., by stimulating users to consider \textit{who} has signed the document and whether the assurances of that individual provide sufficient trust for the document in question. In this regard, a better balance must be struck between providing an easy and fast experience and promoting mindful interactions~\cite{distler2021} with signatures. We will continue exploring solutions to this problem, and we look forward to the HCI community joining us in these endeavors. 

\begin{acks}
We thank the experts who reviewed the \textit{IdentitySign} prototype in the early development phase: Leon Botros, Merel Brandon, Marianna De Sa Siqueira, Arnout Terpstra, and Lian Vervoort. Next, we thank the participants of our usability studies. Furthermore, we thank the CHI'24 reviewers for their thoughtful remarks and Bart Jacobs for his valuable feedback and suggestions.
\end{acks}

\bibliographystyle{ACM-Reference-Format}
\bibliography{digital-dotted-lines}

\appendix

\section{Prototype Screenshots}\label{app:a}

\begin{wrapfigure}{r}{\textwidth}
		\centering
        \includegraphics[width=0.68\textwidth]{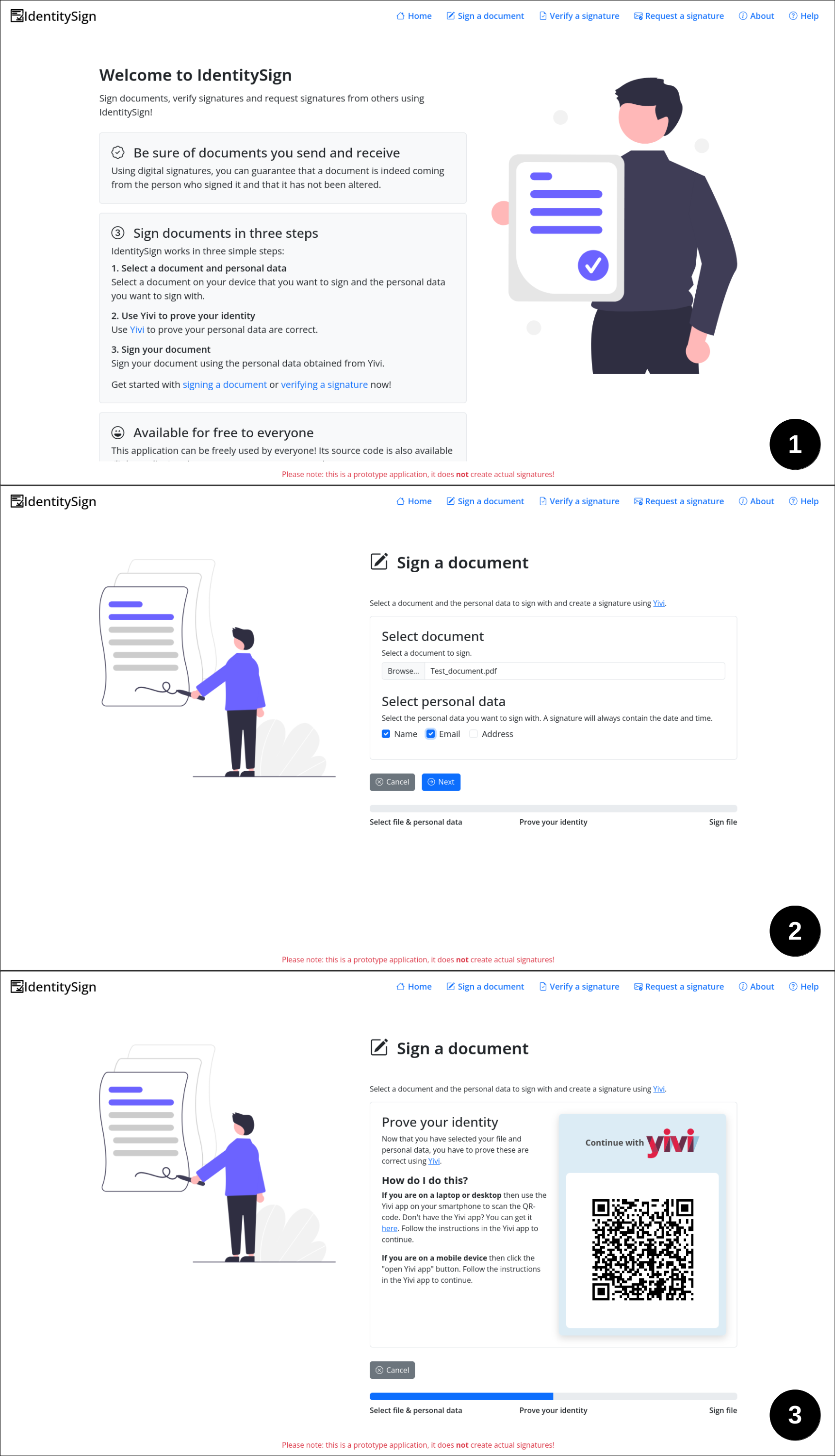}
		\caption{The prototype's home (1) and signing screens (2,3).}
        \Description{A collage of three screenshots of 1: the prototype's home screen, showing a graphic of a person holding a document on the right and text explaining the prototype's functionality and advantages on the left, 2: the prototype's sign screen showing a graphic of a person signing a document on the left and a file selection bar and checkboxes for personal data with name and email checked and a progress bar at the bottom on the right, and 3: another signing screen for the prototype showing the same graphic on the left and a QR-code and instructions for \textit{Yivi} on the right.}
        \label{app:home_sign}
\end{wrapfigure}

\begin{figure*}[ht]
		\centering
        \includegraphics[width=0.68\textwidth]{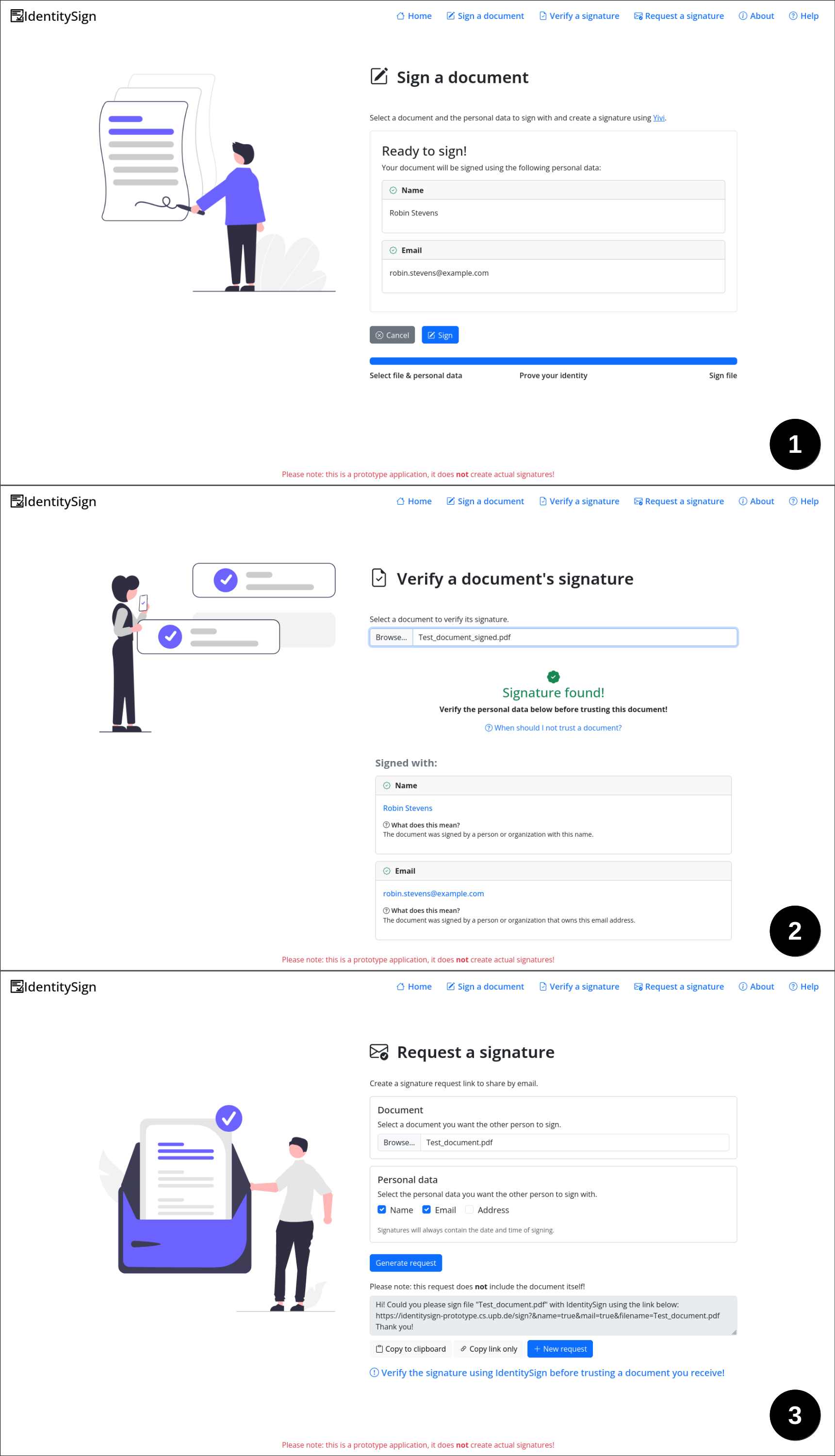}
		\caption{The prototype's signing (1), verifying (2), and requesting (3) screens.}
        \Description{A collage of three screenshots of 1: the prototype's signing screen containing a graphic of a person signing a document on the left half of the screen and a list of cards containing the name and email address for the signature on the right, 2: the prototype's verifying screen containing a graphic of a person and checkmarked fields to the left and a file selection bar with a checkmark and the text "Signature found!" underneath together with cards containing the name and email address used in the signature on the right, and 3: the prototype's request screen showing a graphic of a person holding a document in an envelope to the left and a file selection bar, checkboxes for personal data, a "Generate request" button, and a textbox containing a generated link on the right.}
        \label{app:sign_verify_request}
\end{figure*}

\begin{figure*}[ht]
		\centering
        \includegraphics[width=0.68\textwidth]{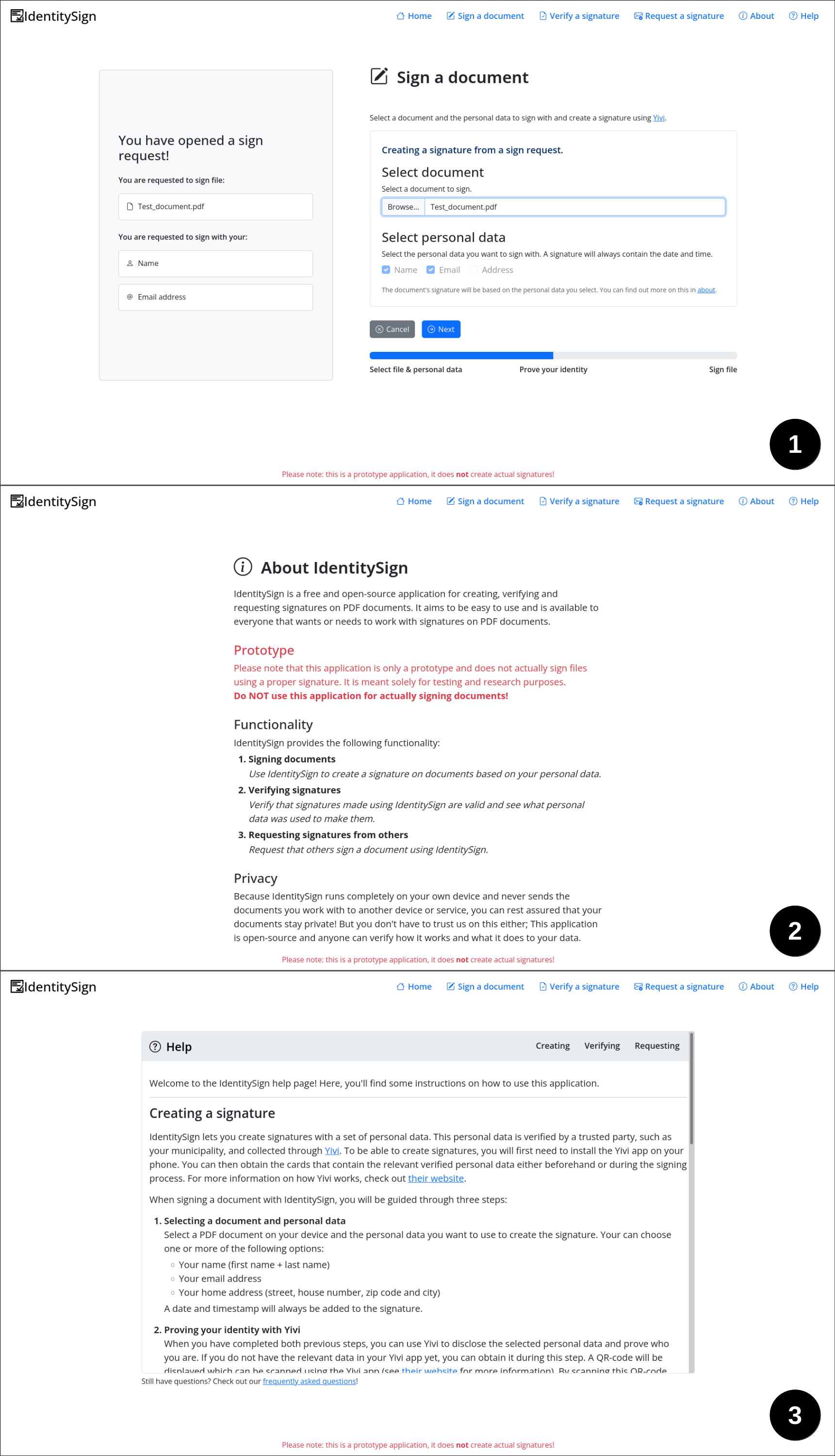}
		\caption{The prototype's requesting (1), about (2), and help (3) screens.}
        \Description{A collage of three screenshots of 1: the signing screen of the prototype when opening a signature request, showing an overview of the filename and personal data from the request on the left half of the screen and a file selection box, pre-checked checkboxes for personal data and a progress bar on the right, 2: the about page of the prototype, containing centered text that explain the prototype and its functionality, and 3: the prototype's help page containing text centered in a scrollable box currently showing (written) instructions on how to create a signature.}
        \label{app:request_about_help}
\end{figure*}

\end{document}